\documentclass[sn-mathphys-num]{sn-jnl}

\usepackage{graphicx}%
\usepackage{multirow}%
\usepackage{amsmath,amssymb,amsfonts}%
\usepackage{amsthm}%
\usepackage{mathrsfs}%
\usepackage[title]{appendix}%
\usepackage{xcolor}%
\usepackage{textcomp}%
\usepackage{manyfoot}%
\usepackage{booktabs}%
\usepackage{algorithm}%
\usepackage{algorithmicx}%
\usepackage{algpseudocode}%
\usepackage{listings}%

\theoremstyle{thmstyleone}%
\theoremstyle{thmstyletwo}%

\theoremstyle{thmstylethree}%

\begin{document}

\title[Threshold resummation]{Threshold resummation for $ZZ$ production}


\author*[1]{\fnm{Pulak Banerjee} \sur{}}\email{pulak.banerjee@lnf.infn.it}


\affil*[1]{
\orgname{Istituto Nazionale di Fisica Nucleare, Gruppo collegato di Cosenza}, \orgaddress{\street{I-87036 Arcavacata di Rende}, \city{Cosenza},  Italy
}}




\abstract{We briefly review the status of threshold resummation for two massive $Z$-bosons in the Standard Model. We discuss some recent results for  $Z$-boson pair production at next-to-next-to-leading order + next-to-next-to-leading logarithmic accuracy. 
}




\maketitle

\section{Introduction}\label{sec1}
Study of massive final state provides a unique opportunity to probe the Standard Model (SM). Some of the interesting aspects being studied over decades include shape of the Higgs potential and electroweak symmetry breaking mechanism. 
Massive vector boson pair production is one of the important electroweak processes at the hadron colliders. It is possible to study the gauge symmetry structure of weak interactions with the help of this process. Production of a vector boson pair is an important background for Higgs production; by measuring the $ZZ$ and $WW$ final states, the off-shell coupling of the Higgs boson can be studied \cite{ATLAS:2014lla}.
Measurement of pair of vector boson in the final state is important for constraining Higgs width \cite{CMS:2021pqj,
ATLAS:2015cuo,
Grazzini:2021iae,
Campbell:2013wga,
Caola:2013yja,
CMS:2014quz}.
Weak bosons decaying into jets,  charged leptons or neutrinos can produce the same signal as new electroweak bosons or any dark matter candidates. The decay of a pair of vector boson to four charged leptons
provide a very clean signal at the collider experiments  \cite{CMS:2021pqj,ATLAS:2015rsx,ATLAS:2016bxw,CMS:2018ccg,CMS:2016ogx,ATLAS:2017bcd,CMS:2017dzg,ATLAS:2019xhj,CMS:2020gtj}.
Electroweak pair production processes are important to place 
constraints on the anomalous trilinear gauge couplings \cite{Hagiwara:1986vm}. 

The production of $ZZ$ in the final state has smaller cross section compared to the other diboson processes.
Precise knowledge of the production cross sections as well as various kinematic distributions at the current LHC and future high energy hadron colliders is necessary.
While the leading-order(LO) corrections were computed decades ago, \cite{Mele:1990bq,Zecher:1994kb,Ohnemus:1994ff} the next-to-leading order(NLO) QCD
corrections for this process can be found in \cite{Campbell:1999ah,Dixon:1998py,Campbell:2011bn}. 
In the context of Beyond Standard Model (BSM) scenarios, 
production of $ ZZ $ at NLO has been studied in \cite{Agarwal:2009xr,Agarwal:2009zg}.
The NLO QCD+Parton-shower(PS) analysis was reported in \cite{Melia:2011tj,Frederix:2011ss,Frederix:2013lga}. 
For an on-shell vector boson pair production, the NLO electroweak(NLO-EW) corrections were presented in
\cite{Bierweiler:2012kw,Bierweiler:2013dja, Gieseke:2014gka}; NLO-QCD+EW corrections for two opposite-charge leptons and two neutrinos in the final state, via decay of off-shell vector bosons have been reported in \cite{Kallweit:2017khh}. For PS effects at NLO-QCD+EW see \cite{Chiesa:2020ttl}.  
The NLO-EW results for inclusive production of $ZZ$,
supplemented with resummed Sudakov logarithms at NLL accuracy has been included in the framework of Sherpa event generator \cite{Bothmann:2021led}.
Study of longitudinally polarized vector bosons can shed light on the electroweak symmetry breaking mechanism; studies at NLO-QCD are available \cite{Hoppe:2023uux}.
A pair of polarized vector bosons at NLO-QCD, matched to PS accuracy has been recently reported in \cite{Pelliccioli:2023zpd}.

Precision studies entail going beyond NLO.
The NLO-QCD corrections to $ZZ$ production in association with jet was done in \cite{Binoth:2009wk}, where the scale uncertainties were found to be at the level of 10\%. The NLO-QCD corrections in association with two jets were studied in \cite{Campanario:2014ioa}. 
To get analytical result for virtual amplitudes at two loop, the master integrals are available \cite{Gehrmann:2013cxs,Gehrmann:2014bfa}, followed by the helicity amplitudes for four lepton production \cite{Gehrmann:2015ora}.
It is also worth noting that the NNLO 
corrections are computed for both on-shell $Z$-boson \cite{Cascioli:2014yka,Heinrich:2017bvg} case as well as for off-shell $Z$-bosons \cite{Grazzini:2015hta,Kallweit:2018nyv}, followed by their 
decay to lepton final states. Fiducial cross sections and distributions are also available
 for the vector boson pair production processes \cite{Grazzini:2015hta,Grazzini:2017ckn}.
The total production cross-section for $ZZ$ has been measured at Large Hadron Collider(LHC) \cite{CMS:2020gtj}; the invariant mass distribution has also been measured up to 1 TeV region. 
The upcoming High Luminosity Large Hadron Collider is expected to reach an integrated luminosity of $3000-4000$ fb$^{-1}$, which can allow the measurement of $ZZ$ events in the TeV region with more statistical data. 
Moreover at the planned Future Circular Collider-hh \cite{FCC:2018vvp}, the increase in parton fluxes along with the integrated luminosity of $20-30$ ab$^{-1}$ can enhance the invariant mass distributions by a few orders of magnitude. It can help to precisely measure the $ZZ$ events beyond $1$ TeV region.
Such accurate measurements must be supplemented with accurate theoretical studies. 
It is important to go beyond NNLO;
efforts have been made through resummation in order to get precise results for various observables. Resummation for transverse momentum at NNLO+NNLL has been achieved in \cite{Grazzini:2015wpa}; at NNLO+N$^3$LL accuracy resummation for the same observable has been presented in  \cite{Campbell:2022uzw}. Using $ \rm MINNLO_{PS} $, the parton shower matched to NNLO results for
different observables can be found in \cite{Buonocore:2021fnj}. An automated framework for resummation of electroweak  Sudakov logarithms using SCET$_{\text{EW}}$ framework has been recently achieved  \cite{Denner:2024yut}.
It is also interesting to consider the gluon fusion channel as well. In the massless quark limit, the NLO corrections for the gluon fusion channel are about $68\%$ of its LO for the current LHC energies \cite{Caola:2015psa}. Inclusion of massive top quark loops increase the corrections to about $73\%$ \cite{Agarwal:2024pod}.
For the same channel, while the LO matched to parton shower are available in \cite{Binoth:2008pr}, the NLO contribution matched to parton shower can found in \cite{Alioli:2016xab}. 

For the invariant mass distribution of $ZZ$ production, threshold resummation at NLO+NNLL accuracy were available long time ago, in \cite{Wang:2014mqt}. At this accuracy, the total cross section increases $\sim$ 7\% for $ZZ$ production. For a pair of gauge bosons, resummation with a
jet-veto at NLO+NNLL accuracy has been studied in 
\cite{Wang:2015mvz}. 
The NNLO+NNLL resummation corrections to invariant mass distribution for $ZZ$ production have been very recently performed in \cite{Banerjee:2024xdh}. Recently an effort has been made to compute three-loop master integrals for the production of two off-shell vector bosons
\cite{Canko:2024ara}.

Our review is structured as following: In the section \ref{sec2}, we shall briefly review the theoretical details; in section \ref{sec3}, we shall present some numerical results. Finally, we conclude in section \ref{sec4}.

\section{Theory}
\label{sec2}
Threshold resummation is an active area of research. At NLL accuracy, resummation for quark initiated processes was achieved via factorization approach in \cite{Sterman:1986aj}. Using an eikonal approach, the authors in
\cite{Catani:1989ne} studied resummation at NLL accuracy, for deep inelastic scattering(DIS) and Drell-Yan(DY) process. Resummation of Sudakov logarithms via factorization approach was presented in  \cite{Contopanagos:1996nh}. 
For a renormalization group approach till NLL accuracy, see \cite{Forte:2002ni}.
It is to be noted that the exponentiation of eikonal cross section in non-Abelian gauge theories was well known for a long time \cite{Gatheral:1983cz,Frenkel:1984pz}.
At the level of scattering amplitudes, exponentiation of soft gluon corrections, using a path integral approach was achieved in \cite{Laenen:2008gt}. There are other approaches to resummation, using Soft-Collinear-Effective Theory(SCET), see \cite{Manohar:2003vb, Pecjak:2005uh, Chay:2005rz, Idilbi:2005ky,Becher:2006mr}. Being an effective theory, SCET 
deals with hadronic degrees of freedom and the resummed results can be derived 
in terms of hadronic kinematic variables \cite{Becher:2006mr}.
In \cite{Bonvini:2012az}, a comparison was made between the QCD approach to resummation and the SCET approach, where the authors have shown that SCET resummation performed in Mellin space(with proper choice of the soft scale) is equivalent to the standard perturbative QCD approach. For more details related to a comparison of the soft gluon resummation in SCET and direct QCD see \cite{Bonvini:2014qga}. 

The perturbative QCD resummation approach, we shall discuss in this review, is from \cite{Catani:1989ne}, and subsequently extended to an all-order resummation result for rapidity distribution of a colorless final state in \cite{Banerjee:2017cfc}. This approach was later used for rapidity distribution for DY at NNLO+NNLL \cite{Banerjee:2018vvb},  Higgs boson production via bottom quark annihilation at N$^3$LO +  N$^3$LL \cite{AH:2019phz}, DY at N$^3$LL \cite{Ajjath:2020rci}, for colorless production in the final state \cite{Das:2022zie}; for $ZH$ production \cite{Das:2025wbj}. For rapidity distribution of $Z, W^{\pm}$ beyond NNLL see \cite{Das:2023bfi}, and for the same observable for Higgs boson production via bottom quark annihilation see \cite{Das:2023rif}.
Application at beyond the soft approximation can be found in \cite{AH:2020iki,AH:2022lpp,AH:2021kvg,AH:2021vdc,AH:2020qoa,AH:2021vhf,Ravindran:2022aqr,Ahmed:2020amh,AH:2020xll,Bhattacharya:2021hae,Bhattacharya:2025rqk}.
For different BSM models the resummed cross section has been presented in \cite{Agarwal:2018vus,Das:2019bxi,Das:2020gie,Ravindran:2023qae,Ahmed:2016otz,Das:2020pzo}. The formalism has been extended to $n$ colorless particle in the final state in \cite{Ahmed:2020nci} and subsequently used to predict the N$^3$LO + N$^3$LL cross section  for a pair of Higgs in the final state \cite{AH:2022elh}. We have recently applied the formalism for a pair of $Z$ bosons, produced via quark-antiquark annihilation in \cite{Banerjee:2024xdh}, and this will be the topic of discussion for the rest of the review.
It is to be noted that there exists other approaches to threshold resummation for rapidity distributions in DY process, a detailed comparison in between these approaches has been made in \cite{Bonvini:2023mfj}.

In the rest of the section, we briefly review the theory; for a detailed discussion see \cite{Ravindran:2005vv,Ravindran:2006cg, Ahmed:2020nci,Banerjee:2017cfc}. It is well known that collinear factorization allows to express the hadronic cross section in terms of the partonic counterpart as:
\begin{align}\label{eq:had-xsect}
 \frac{d\,\sigma}{d\, Q}  = &
\sum_{a,b= \{q, \bar{q}, g\}}\int_0^1 dx_1\int_0^1 dx_2 f_{a}(x_1,\mu_F^2)
f_{b}(x_2,\mu_F^2)   
 \int_0^1 dz~ \delta \left(\tau- z\, x_1 x_2 \right) 
\frac{d\,\hat\sigma_{ab}}{d\, Q}\,,
\end{align}
where the threshold variables are $\tau=\frac{Q^2}{S}$ and $z= \frac{Q^2}{s}$.
$S$ and $s$ are the hadronic and partonic
center of mass energies respectively; $\mu_F$ is the mass factorization scale.
The partonic cross section, $\frac{d\,\hat\sigma_{ab}}{d\, Q}$, can be expanded in a perturbative series of the strong coupling constant.
Beyond LO, $\frac{d\,\hat\sigma_{ab}}{d\, Q}$ develops infrared and ultraviolet divergences. Physical cross-sections are inclusive over arbitrary soft particles produced in the final state. The infrared divergent virtual gluons cancel the infrared divergences from undetected real gluons. Collinear factorization allows us to absorb the initial state collinear singularities in the mass factorization kernels.
After the cancellation of infrared and ultraviolet divergences, the finite cross-section contains plus distributions ${\mathcal {D}_{i}} = \left(\frac{\log^i[1-z]}{1-z}\right)_{+}$, delta functions ($\delta(1-z)$) and regular terms.  A particularly interesting limit is $z\rightarrow1$, which is the threshold limit, where the final state ($ZZ$ in our discussion with $a=q, b= \bar{q}$) carries most of the total energy. This leads to suppression of the radiative tail of the real emission processes and causes terms like $a_s {\mathcal {D}_{i}}$ $\gtrsim 1$. The sv(soft virtual) contributions, consisting of plus distributions and delta functions give dominant contributions in this limit in comparison to the regular terms. 
These sv contributions needs to be resummed to all orders in perturbation theory. 

The sv contributions of the partonic cross section can be expressed in a  factorized form in  terms of hard form factor, mass factorization kernel
and soft-collinear radiations, see \cite{Ravindran:2005vv,Ravindran:2006cg} for details.
It is to be noted that in \cite{Ravindran:2005vv,Ravindran:2006cg} a $z$-space formalism was developed
to obtain soft distribution function that can be used to obtain 
threshold enhanced cross section for the inclusive production of any colorless particle. The author in \cite{Ravindran:2006cg} has shown that the $N$-th Mellin moment of finite part of the universal soft distribution function was the threshold exponent \cite{Sterman:1986aj,Catani:1989ne,Catani:1990rp}. Using the same approach, the DY production at threshold at N$^3$LO was obtained in \cite{Ahmed:2014cla},  rapidity distributions was obtained for 
lepton pairs, Higgs boson \cite{Ravindran:2006bu} 
and $Z$ and $W^{\pm}$ \cite{Ravindran:2007sv} in the threshold limit up to N$^3$LO level \cite{Ahmed:2014uya,Ahmed:2014era}.  Partial sv results at N$^4$LO  QCD  are available for inclusive Higgs production \cite{Das:2020adl}. While the threshold enhanced corrections to inclusive Higgs production through bottom anti-bottom annihilation at N$^3$LO was studied in \cite{Ahmed:2014cha}, threshold enhanced corrections for inclusive DIS at 4 loops can be found in \cite{Das:2019btv}.
For sv contributions to the bottom quark energy distribution in Higgs boson decay at NNLO see \cite{Blumlein:2006pj}. 
One can construct a soft-collinear operator for threshold resummation which contains terms proportional to both  delta functions and ${\mathcal {D}_{i}}$. 
For a discussion about how the sv cross section is related to the universal soft-collinear operator, see \cite{Ahmed:2020nci}. In Mellin space the sv cross section 
has the following exponential form:
\begin{align}
\label{eq:res1}
\int_0^1 dz \, z^{N-1} \, \frac{d\,\hat\sigma_{q\bar{q}}^{\text{sv}}}{d\, Q} 
&= \overline {g}_{0}^{q\bar{q}} \exp\Big( \overline G_{\bar{N}}^{q\bar{q}}\left(\omega\right)\Big)\
\end{align}
where 
\begin{align}
\label{eq:GNbexp}
 \overline g_{0}^{\,q\bar{q}} & =  \nonumber \sum_{i=0}^\infty a_s^i(\mu_R^2) \overline g_{0,i}^{\,q\bar{q}}\\
 {\overline G}_{\bar{N}}^{\,q\bar{q}}(q^2,\mu_F^2,\omega) & = \ln \bar{N}\, {\overline g}_{1}^{\,q\bar{q}}(\omega) + \sum_{i=0}^\infty a_s^i(\mu_R^2){\overline g}_{i+2}^{\,q\bar{q}}(\omega,q^2,\mu_F^2, \mu_R^2) .
\end{align}
Here $\omega= 2\beta_0 a_s(\mu_R^2) \ln \bar{N}$,  $\bar{N} = N \exp(\gamma_E)$ and $\gamma_E$ is the Euler-Mascheroni constant.
The resummation exponents($g_{i}^I$)($I\in\{gg, q \bar{q}\}$), needed for resummation till N$^3$LL accuracy can be found in
\cite{Banerjee:2017cfc, Banerjee:2018vvb, Ahmed:2020nci}. For a brief description of how these resummed coefficients are calculated, see Appendix 2 of \cite{Banerjee:2018fnt}.
These depend on the nature of the external particles, quarks or gluons.
At NNLO+NNLL accuracy, the ${\mathcal O(\alpha_s^2)}$ virtual matrix elements are included in $g_{0,2}^{\,q\bar{q}}$. It is to be noted that while the Mellin transformation of ${\mathcal {D}_{i}}$ are included only inside the exponential(see Eq. 12 of \cite{Banerjee:2017cfc}), all the $\delta(1-z)$ contributions(from virtual contributions as well as soft-collinear operator) has been included inside $g_{0,i}^{\,q\bar{q}}$. It is also possible to resum part (or full)
of the $g_0^{I}$ by including them in the exponent
\cite{Bonvini:2014joa,Bonvini:2016frm,Eynck:2003fn,Das:2019btv,Ajjath:2020rci},
which however have sub-leading effects. Finally the matched results, necessary for the phenomenological applications, can be obtained
via inverse Mellin transformation.
The Mellin inversion can then be performed \cite{Vogt:2004ns} by choosing a proper contour.
The issue of Landau pole during Mellin inversion has been well known \cite{Amati:1980ch} and we follow the minimal prescription scheme \cite{Catani:1996yz}.
In the next section we briefly discuss some results for on-shell $ZZ$ production at NNLO+NNLL.

\section{$ZZ$ production at NNLO+NNLL}
\label{sec3}
\begin{figure}[ht!]
	\centerline{
		\includegraphics[scale =0.35]{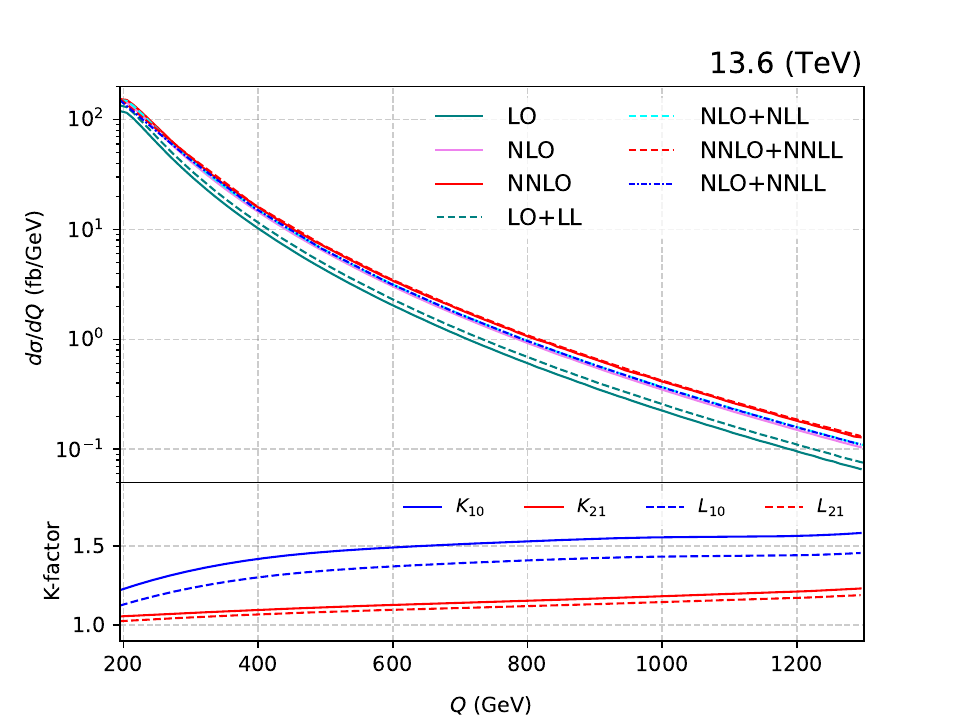}
		\includegraphics[scale =0.35]{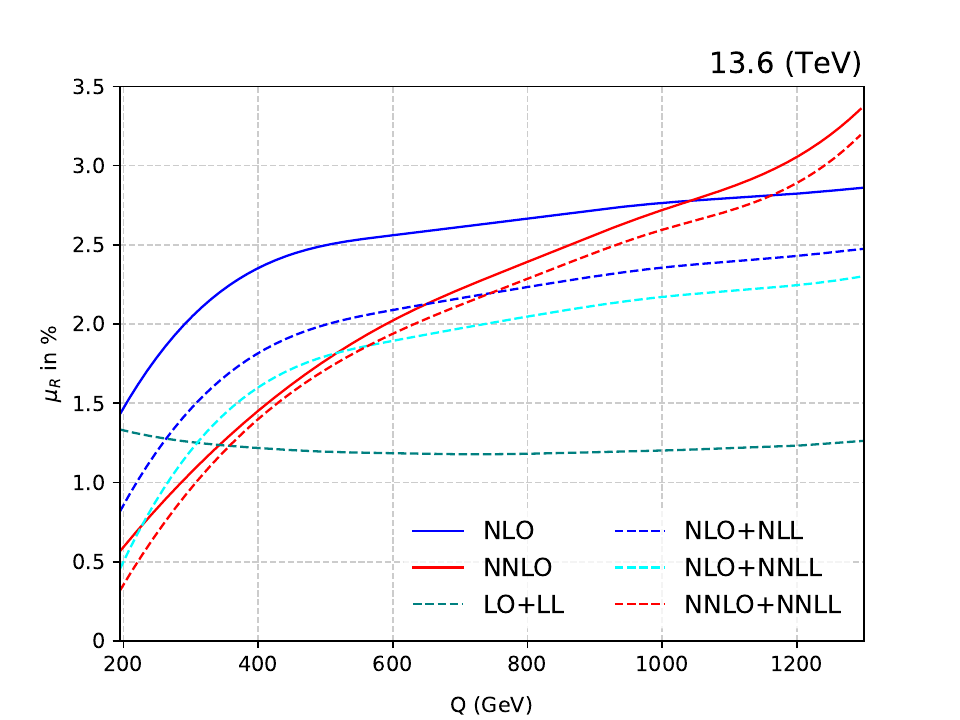}
	}
	\vspace{-2mm}
	\caption{\small{Resummed invariant mass distribution along with the k-factors(left) and $\mu_R$ scale uncertainties (right)  for  $Z$-boson pair production up to NNLO+NNLL.}}
	\label{fig:invmass_scale}
\end{figure}

We present a brief discussion on the numerical implication for $ZZ$ production at NNLO+NNLL, for more details see \cite{Banerjee:2024xdh}. Being an important observable, precise predictions are necessary and the resummation accuracy should match the same accuracy as the available fixed order. 
The fixed order results have been obtained from MATRIX \cite{Grazzini:2017mhc}, while the resummed cross-section are obtained from our in-house codes. 
In left hand side of Fig. \ref{fig:invmass_scale} we plot the
invariant mass distribution along with the
K-factors($K_{\text{ij}}=\frac{\sigma_{\text{N}^i\text{LO}}}{\sigma_{\text{N}^j\text{LO}}}$, $L_{\text{ij}}=
\frac{\sigma_{\text{N}^i\text{LO} + \text{N}^i\text{LL}}}{\sigma_{\text{N}^j\text{LO}+\text{N}^j\text{LL}}}$).
For $Q=$ 1300 GeV, the NLO corrections are about 58\% of the LO, the NNLO corrections are about 94\% of the LO. For the same Q, the NLO+NLL results are about 68\% of the LO, while the NNLO+NNLL results are about 99\% of the LO.
It is to be noted that the NLL logarithms that are not present in LO+LL, are about 45\% of the LO+LL. The NNLL logarithms contribute an additional 18\% over NLO+NLL, which is non-negligible. 
It is interesting to consider the scenario when NNLL corrections are added on the top of NLO instead of NNLO, as was done in \cite{Wang:2014mqt}. We study it in the right hand side plot of
Fig. \ref{fig:invmass_scale}, where we have shown the  renormalization scale uncertainties, which are estimated by fixing $\mu_F$ and varying $\mu_R$ in the range: $\{Q/2 \le \mu_R \le 2 Q\}$.
For the NLO+NNLL plot, we have generated the NLO result with MSHT20nlo parton distribution functions (PDFs) \cite{Bailey:2020ooq} and NNLL with MSHT20nnlo PDFs. The renormalization scale uncertainty at NLO+NNLL improves in comparison to NLO+NLL, as NNLL adds higher-order logarithms.
On the other hand, the NNLO K-factor
slowly but continuously increases(1.29 to 1.94), as can be seen in Fig 1 of \cite{Banerjee:2024xdh}. This clearly shows additional contributions 
coming from second-order corrections in the higher $Q$-region, which are due to the real correction sub-processes like $qq^\prime \to ZZ qq^\prime$. Thus it is important to 
consider these additional contributions while doing the resummation at NNLL accuracy. We also note that the seven point scale uncertainty plot in Fig 3 of \cite{Banerjee:2024xdh} shows the improvement in scale uncertainties at NNLO+NNLL, in comparison to NNLO.
We also observe from the left plot of Fig \ref{fig:invmass_scale}, that the NLO+NNLL invariant mass distribution is closer to NLO+NLL. The invariant mass distribution at NLO+NNLL deviates from NNLO+NNLL by about 3\%  to 16\%, for the entire $Q$ range.  

In table 1, we present the invariant mass distribution at NNLO+NNLL for $\mu_R=\mu_F=Q$, for different PDFs, namely ABMP16\cite{Alekhin:2016uxn}, CT18\cite{Hou:2019efy}, PDF4LHC21\cite{PDF4LHCWorkingGroup:2022cjn} and MSHT20. We observe that the invariant mass distribution for each PDF coincides with each other, at different $Q$ values, within their errors.
\begin{table}[ht!]
\label{tab:pdftab}%
\begin{tabular}{@{}lllll@{}}
\toprule
Q in GeV & \quad MSHT & \quad PDF4LHC & \quad ABMP & \quad CT\\
\midrule
\vspace{0.2cm}
295    & 48.294\,$^{+0.578}_{-0.569}$   & 48.118\,$^{+0.566}_{-0.559}$   &
47.433 \,$^{+0.535}_{-0.527}$  & 
47.668\,$^{+0.557}_{-0.552}$   \\
\vspace{0.2cm}
595    & \,\,\,\,\,\,3.651\,$^{+0.099}_{-0.086}$ &  
 \,\,\,\,\,\,3.584\,$^{+0.077}_{-0.071}$   &
 \,\,\,\,\,\,3.514 \,$^{+0.067}_{-0.062}$    & \,\,\,\,\,\,3.524\,$^{+0.071}_{-0.066}$  \\
995    & \,\,\,\,\,\,0.431\,$^{+0.014}_{-0.013}$   & 
\,\,\,\,\,\,0.428\,$^{+0.014}_{-0.013}$   &
\,\,\,\,\,\,0.422\,$^{+0.013}_{-0.012}$  & \,\,\,\,\,\,0.423\,$^{+0.014}_{-0.013}$  \\
\botrule
\caption{Invariant mass distribution(in fb/GeV) for $\mu_R=\mu_F=Q$ at NNLO+NNLL for different PDFs.}
\end{tabular}
\end{table}



\section{Conclusion}
\label{sec4}
In this review, we briefly discussed the resummation of soft gluon contributions, at threshold, for a pair of $Z$ bosons produced via quark-antiquark annihilation.
After discussing about the current accuracy of $ZZ$ production both in fixed-order as well as in resummation, we briefly mention, in general, about the different approaches to threshold resummation. Finally we discuss the numerical implication of the resummation at NNLO+NNLL accuracy, for a pair of on-shell $Z$ bosons.

\bmhead{Acknowledgments}
We thank A.H.Ajjath, C. Dey, M.C. Kumar, V. Pandey, A. Papa and V. Ravindran for useful discussions.


\begin{appendices}




\end{appendices}


\bibliography{main}

 \,.

\end{document}